\documentclass[reprint,showpacs,showkeys,preprintnumbers,amsmath,amssymb,aip,apl]{revtex4-1}

\usepackage{graphicx}
\usepackage{float}
\usepackage{color}

\begin{document}
\title{A novel method to evaluate spin diffusion length of Pt}
\author{Y. Q. Zhang}
\affiliation{Shanghai Key Laboratory of Special Artificial Microstructure and Pohl Institute of Solid State Physics and School of Physics Science and Engineering, Tongji University, Shanghai 200092, China}
\author{N. Y. Sun}
\affiliation{Shanghai Key Laboratory of Special Artificial Microstructure and Pohl Institute of Solid State Physics and School of Physics Science and Engineering, Tongji University, Shanghai 200092, China}
\author{W. R. Che}
\affiliation{Shanghai Key Laboratory of Special Artificial Microstructure and Pohl Institute of Solid State Physics and School of Physics Science and Engineering, Tongji University, Shanghai 200092, China}
\author{R. Shan}\email{shan.rong@hotmail.com}
\affiliation{Shanghai Key Laboratory of Special Artificial Microstructure and Pohl Institute of Solid State Physics and School of Physics Science and Engineering, Tongji University, Shanghai 200092, China}
\author{Z. G. Zhu}\email{zgzhu@ucas.ac.cn}
\affiliation{School of Electronic, Electrical and Communication Engineering, University of Chinese Academy of Sciences, Beijing 100049, China.}
\affiliation{Theoretical Condensed Matter Physics and Computational Materials Physics Laboratory, College of Physical Sciences, University of Chinese Academy of Sciences, Beijing 100049, China.}

\date{\today}

\begin{abstract}

Spin diffusion length of Pt is evaluated via proximity effect of spin orbit coupling (SOC) and anomalous Hall effect (AHE) in Pt/Co$_{2}$FeAl bilayers. By varying the thicknesses of Pt and Co$_{2}$FeAl layer, the thickness dependences of AHE parameters can be obtained, which are theoretically predicted to be proportional to the square of the SOC strength. According to the physical image of the SOC proximity effect, the spin diffusion length of Pt can easily be identified from these thickness dependences. This work provides a novel method to evaluate spin diffusion length in a material with a small value.

\end{abstract}


\maketitle
%\section{Introduction} %%%%%%%%%%%%%%%%%%%%%%%%%%%%%%%%%%%%%%%%%
As a promptly growing research area, spintronics aims at using and manipulating not only the charge, but also the spin in an electronic device~\cite{Wolf2001,I2004}. Spin dependent transport property is the key for the application of spintronics device, and thus it has attracted great attention over the past few decades~\cite{G1988,Nakayama2013,lin2014,Althammer2013,chen2013}. Herein, spin diffusion length ($\lambda$) is a fundamental parameter in the study of spin dependent transport, which has an inverse relationship with the intensity of spin dependent scattering. There are many methods such as lateral spin valve, spin pumping, spin-torque ferromagnetic resonance, Hall cross and spin absorption to gain the spin diffusion length in a normal metal~\cite{Laczkowski2012,Vlaminck2013,Zhang2013,Azevedo2011,JC2014,liu,G2009,Isasa2015}. Different from above methods, a novel approach to evaluate spin diffusion length is introduced in this work, through proximity effect of spin orbit coupling (SOC) and anomalous Hall effect (AHE)~\cite{Hall1881,Nagaosa2010,zhang15}.
%Figure 1%%%%%%%%%%%%%%%%%%%%%%%%%%%%%%%%%%%%%%%%%%%%%%%%%%%%%%%%%%%%%%%%%%%%%%%%%%%%%%%%%
\begin{figure}[t]
\centering
   \includegraphics[width=6.5cm]{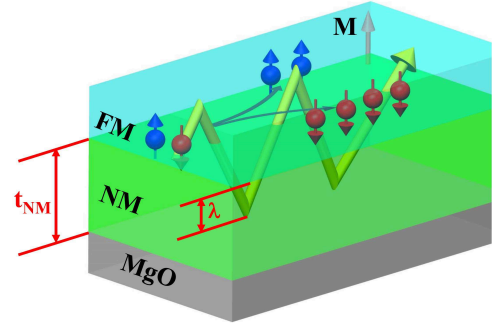}
   \caption{(Color online) Physical image of AHE in FM/NM bilayer. Gray arrow (M) indicates the magnetic moment direction of FM layer. Yellow line indicates possible motion path of the conduction electrons. $t_{NM}$ and $\lambda$ is thickness and spin diffusion length of NM layer, respectively.}
\label{Fig1}
\end{figure}
%Figure 1%%%%%%%%%%%%%%%%%%%%%%%%%%%%%%%%%%%%%%%%%%%%%%%%%%%%%%%%%%%%%%%%%%%%%%%%%%%%%%%%%
It is well known that spin orbit coupling plays a fundamental role in spin dependent transport properties, including anomalous Hall effect, spin Hall effect, spin transfer torque and Dzyaloshinskii-Moriya interaction etc.~\cite{Hall1881,Sinova2004,Sloncz1996,Dzy1958,Mor1960}. SOC can strongly affect the intensity of spin dependent scattering, and hence has a direct relationship with spin diffusion length. If a low dimensional layered structure consists of a nonmagnetic (NM) layer with strong SOC and a ferromagnetic (FM) layer, the conduction electrons will be repeatedly scattered in z direction by the interfaces of heterolayers, substrates and vacuum when the current is flowing, as shown in Fig.~\ref{Fig1}. The conduction electrons will be polarized by the magnetic layer and affected by the strong SOC layer meantime, causing enhanced spin dependent scattering. This phenomenon is called SOC proximity effect. When the thickness of the NM layer is over its spin diffusion length, the added NM layer will not contribute to the SOC proximity effect but continue shunting the measurement current, which leads to a  suddenly weaken performance of spin dependent scattering. Catching the turning point, the spin diffusion length can consequently be evaluated.

In order to analyze spin dependent scattering, anomalous Hall effect is employed in this work. It is now firmly established that there are two categories competing mechanisms contributing to the AHE: intrinsic mechanism, which originates from the anomalous velocity of the Bloch electrons induced by the SOC~\cite{Karplus1954,Jung2002,Xiao2010}; external mechanism, which includes skew scattering and side jump~\cite{Smi55,Ber70}. Both intrinsic and side jump contribution obey the square relationship $\rho_{AH}\propto\rho^2_{xx}$, where $\rho_{AH}$ and $\rho_{xx}$ correspond to the anomalous Hall resistivity and longitudinal resistivity, respectively~\cite{Karplus1954,Jung2002,Xiao2010,Smi55}. Differently, due to asymmetric scattering from impurities caused by SOC, the skew scattering contribution shows linear dependence on $\rho_{xx}$~\cite{Ber70}. Therefore, the measured $\rho_{AH}$ usually obeys the empirical Eq.~(\ref{PAH}):
\begin{equation}
\rho_{AH} = a\rho_{xx} + b\rho^2_{xx},
\label{PAH}
\end{equation}
%\indent
where $a$ and $b$ are parameters. This relationship is known as the traditional scaling. In 2001, Cr\'{e}pieux and Bruno presented a theory of the AHE, where they deem that no matter skew scattering or side jump contributions, AHE parameters ($a$ and $b$) always obey square relationships with strength of SOC~\cite{CB theory}. Afterward, Tian \textit{et al}. proposed an expanded scaling~\cite{Tia09}, where the impurity and phonon are assumed to have different contributions to the skew scattering. The scaling can be rewritten as Eq.~(\ref{PAH1}):
\begin{eqnarray}
\rho_{AH} &=& a'\rho_{xx0} + a''\rho_{xxT} + b\rho^2_{xx}
\label{PAH1}
\end{eqnarray}
Here, $\rho_{xx0}$ is the residual resistivity, $\rho_{xxT}$ comes from the scattering of excited phonons, $a'$ and $a''$ are due to the skew scattering, and $b$ is dominated by the side jump and intrinsic contributions. In this equation, the phonon contribution is introduced by considering the inelastic scattering at finite temperatures, which was formulated in terms of a multiband tight-binding model by Shitade and Nagaosa~\cite{Shitade2012}.\\

%\section{Experimental details}
\indent
Two series of Pt (2.5 nm)/Co$_{2}$FeAl ($t_{CFA}$ nm) and Pt ($t_{Pt}$ nm)/Co$_{2}$FeAl (0.9 nm) bilayer films were deposited on the $1\times 1$ cm$^2$ polished MgO(100) substrates by magnetron sputtering. All those samples were prepared under Hall bar mask and annealed at 320 $^\circ$C in situ. The base pressure of sputtering chamber is below $3\times 10^{-6}$ Pa. The sputtering Ar gas with 99.999$\%$ purity was introduced with a constant pressure of 0.3 Pa. Film thickness was measured by X-ray reflectivity (XRR) using a D8 Discover X-ray diffractometer. The transport property was measured by physical property measurement system (PPMS) from 20 to 300 K.\\
% Figure 2 %%%%%%%%%%%%%%%%%%%%%%%%%%%%%%%%%%%%%%%%%%%%%%%%%%%%%%%%%
\begin{figure}
\centering
   \includegraphics[width=0.75\linewidth]{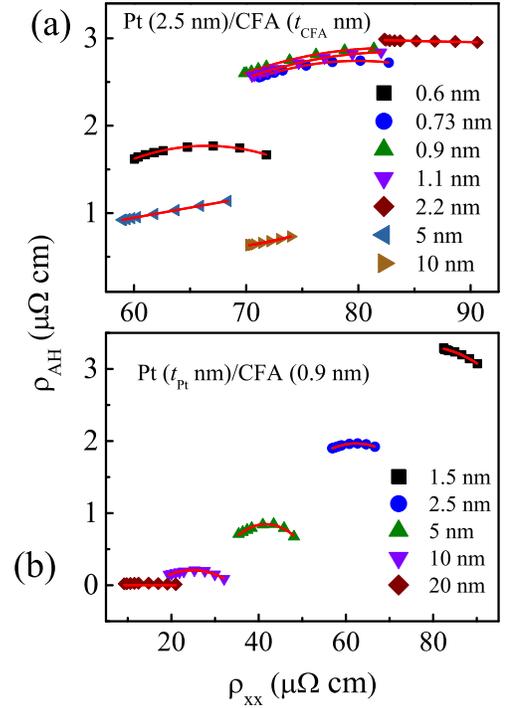}
   \caption{(Color online) $\rho_{AH}/\rho_{xx}$ versus $\rho_{xx}$ curves for (a) Pt (2.5 nm)/Co$_{2}$FeAl (t$_{CFA}$ nm) and (b) Pt (t$_{Pt}$ nm)/Co$_{2}$FeAl (0.9 nm) bilayers. All scattered symbols represent experimental data. All red lines were fitted results by $\rho_{AH}=a'\rho_{xx0}+a''\rho_{xxT}+b\rho^2_{xx}$.}
\label{fig:AHE}
\end{figure}
%%%%%%%%%%%%%%%%%%%%%%%%%%%%%%%%%%%%%%%%%%%%%%%%%%%%%%%%%%%%%%%%%%%%
%\indent

%\section{Results and discussion}
Figure~\ref{fig:AHE} shows the measured results of the AHE. $\rho_{AH}/\rho_{xx}$ versus $\rho_{xx}$ curves of the samples are given in Fig.~\ref{fig:AHE}(a) and (b). The red lines are fitting curves. Obviously, $\rho_{AH}/\rho_{xx}$ versus $\rho_{xx}$ is not linear relationship for almost all samples. Just when Co$_{2}$FeAl (CFA) layer is thicker than 5 nm, the relationship looks like linear. Hence, the traditional scaling does not work well for very thin bilayers. On the contrary, Eq.~(\ref{PAH1}) can fit the data perfectly. These results are consistent with our reported data~\cite{Che2014,q2013}. It reveals that the dependence of $\rho_{AH}/\rho_{xx}$ versus $\rho_{xx}$ shows a straighter line at special ratios among $a'$, $a''$ and b$\rho_{xx0}$ only, such as the cases in a bulk material and a thick film.\\
\indent
According to the CB theory and the physical image of the SOC proximity effect, the effective SOC strength of Pt/Co$_{2}$FeAl can be treated using tight-binding sense, and the expression can be read as:
\begin{equation}
\zeta_{\text{eff}} = \frac{\zeta_{\text{CFA}}t_{\text{CFA}}+\zeta_{\text{Pt}}t_{\text{Pt}}}{t_{\text{CFA}}+t_{\text{Pt}}},
\label{zetaeff}
\end{equation}
where $\zeta_{\text{CFA}} (= 53.8$ meV) represents the SOC strength of Co$_{2}$FeAl, $\zeta_{\text{Pt}} (= 554$ meV) is the SOC strength of Pt, and $t_{\text{CFA(Pt)}}$ is the thickness of Co$_{2}$FeAl(Pt) layer~\cite{kota}.%\\
%\indent

\begin{figure}
\centering
   \includegraphics[width=1\linewidth]{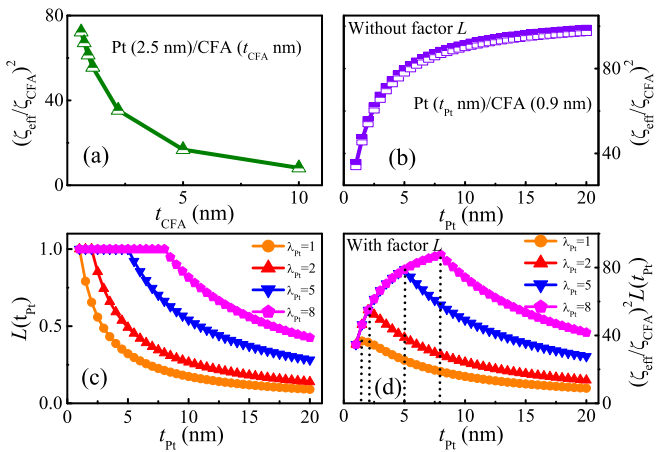}
   \caption{(Color online) Thickness dependence of $(\zeta_{\text{eff}}/\zeta_{\text{CFA}})^2$ given by Eq.~(\ref{zetaeff}) for (a) Pt (2.5 nm)/Co$_{2}$FeAl ($t_{CFA}$ nm) and (b) Pt ($t_{Pt}$ nm)/Co$_{2}$FeAl (0.9 nm) bilayers. (c) Thickness dependence of $L(t_{\text{Pt}})$ for Pt ($t_{Pt}$ nm)/Co$_{2}$FeAl (0.9 nm) bilayer. $L(t_{\text{Pt}})$ is a factor resulting from the leakage current for a thicker Pt layer. $L(t_{\text{Pt}})=1$ when $t_{\text{Pt}}$ $\leq$ $\lambda_{\text{Pt}}$; simply $L(t_{\text{Pt}})=(\lambda_{\text{Pt}}+t_{\text{CFA}})/(t_{\text{Pt}}+t_{\text{CFA}})$ when $t_{\text{Pt}}$ $>$ $\lambda_{\text{Pt}}$. (d) Thickness dependence of $(\zeta_{\text{eff}}/\zeta_{\text{CFA}})^2L(t_{\text{Pt}})$ for Pt ($t_{\text{Pt}}$ nm)/Co$_{2}$FeAl (0.9 nm) bilayer, corresponding to the values with the leakage current factor revision. $\lambda_{\text{Pt}}$ is Pt diffusion length, which is arbitrarily taken as 1, 2, 5, 8 nm in (c) and (d).}
\label{fig:SDL}
\end{figure}

If $t_{\text{Pt}}<\lambda_{\text{Pt}}$ ($\lambda_{\text{Pt}}$ is the spin diffusion length of Pt), increasing the thickness of Pt layer will enhance the effective SOC (which has the meaning of squre of $\zeta_{\text{eff}}$ hereafter) monotonically obeying Eq. (\ref{zetaeff}) shown in Fig.~\ref{fig:SDL}(b). On the contrary, increasing the thickness of the Co$_{2}$FeAl layer will not affect the spin diffusion in the Pt layer but reduce the effective SOC since the atomic SOC in the Co$_{2}$FeAl layer is much weaker than that in Pt layer, which is shown in Fig.~\ref{fig:SDL}(a).

% Figure 3 %%%%%%%%%%%%%%%%%%%%%%%%%%%%%%%%%%%%%%%%%%%%%%%%%%%%%%%%%
%%%%%%%%%%%%%%%%%%%%%%%%%%%%%%%%%%%%%%%%%%%%%%%%%%%%%%%%%%%%%%%%%%%%

If $t_{\text{Pt}} > \lambda_{\text{Pt}}$, the spin states will be lost in partial of Pt layer where the distance to the interface of the FM and NM layer is over the $\lambda_{\text{Pt}}$ already. The reason is that the injected spin (along the $\mathbf{M}$ direction) from the FM layer into Pt layer will be relaxed due to the SOC of Pt. Therefore, this part of Pt only plays a role of conducting a leakage current and a leakage current factor has to be introduced, which is simply proposed as $L(t_{\text{Pt}})=(\lambda_{\text{Pt}}+t_{\text{CFA}})/(t_{\text{Pt}}+t_{\text{CFA}})$ here. Only those Pt atomic layers in the scope of $\lambda_{\text{Pt}}$ can make contribution to the measured signal of spin dependent scattering, e.g. the AHE. The general behavior of factor $L(t_{\text{Pt}})$ is shown in Fig.~\ref{fig:SDL}(c). Taking into account the $L(t_{\text{Pt}})$ factor, the effective SOC of the bilayer system should take behavior shown in Fig.~\ref{fig:SDL}(d), in which the turning point occurs exactly at $t_{\text{Pt}}=\lambda_{\text{Pt}}$. Since the fitted parameters of the AHE are theoretically proportional to the effective SOC of the bilayer system, identifying the turning points in the fitted parameters of the AHE provides a powerful tool to fix the spin diffusion length in the Pt layer.

\begin{figure}
\centering
   \includegraphics[width=1\linewidth]{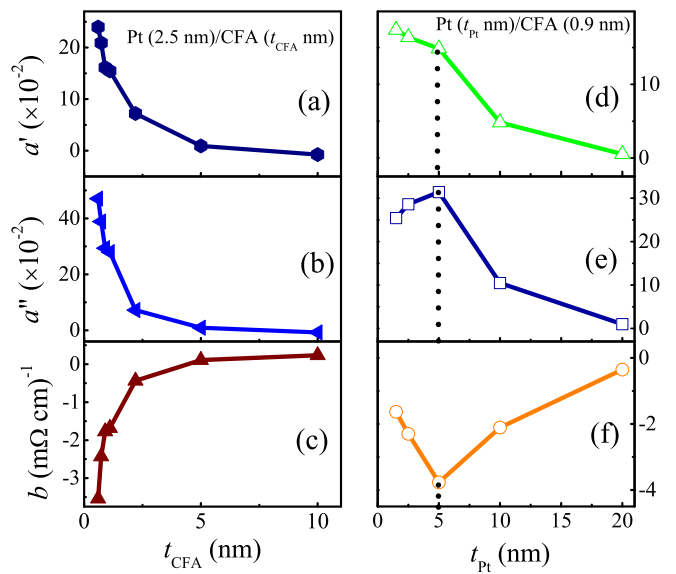}
   \caption{(Color online) Thickness dependences of $a'$, $a''$ and $b$ for Pt (2.5 nm)/Co$_{2}$FeAl ($t_{\text{CFA}}$ nm) in (a)-(c) and Pt ($t_{\text{Pt}}$ nm)/Co$_{2}$FeAl (0.9 nm) bilayers in (d)-(f). Here, $a'$ and $a''$ indicate the contributions of skew scattering from impurities and phonons; $b$ represents the intrinsic and side jump contributions.}
   \label{Fig 4}
\end{figure}

Fig.~\ref{Fig 4} shows the thickness dependences of fitted parameters from Fig.~\ref{fig:AHE}(a) and (b). For the samples with $t_{\text{Pt}}=2.5$ nm, the $t_{\text{CFA}}$ dependences are consistent with those in Fig.~\ref{fig:SDL}(a), indicating $t_{\text{Pt}} < \lambda_{\text{Pt}}$. For the samples with varied thicknesses of Pt layers but fixed thickness of Co$_{2}$FeAl layers, the overall behavior of $a'$, $a''$ and $b$
are found to be similar to those of theoretical prediction shown in Fig.~\ref{fig:SDL}(d) including the leakage current factor. Thus $\lambda_{\text{Pt}}$ is evaluated around 5 nm which is approximate to that in Ref. [\onlinecite{Mar2014}].\\
%Figure 4%%%%%%%%%%%%%%%%%%%%%%%%%%%%%%%%%%%%%%%%%%%%%%%%%%%%%%%%%%%%%%%%%%%%%%%%%%%%%%%%%
%Figure 4%%%%%%%%%%%%%%%%%%%%%%%%%%%%%%%%%%%%%%%%%%%%%%%%%%%%%%%%%%%%%%%%%%%%%%%%%%%%%%%%%

%\section{Conclusions}
\indent
In conclusion, this work demonstrates a method for determining the spin diffusion length via anomalous Hall effect and proximity effect of spin orbit coupling. We obtain the spin diffusion length of Pt is around 5 nm. This method can be used to evaluate nonmagnetic materials with short spin diffusion length. However, due to the limit of proximity effect, it does not work for thick films probably.

%Acknowledgement %%%%%%%%%%%%%%%%%%%%%%%%%%%%%%%%%%%%%%%%%%%%%%%%%%%%%
\begin{acknowledgments}
We thank the beamline 08U1 at the Shanghai Synchrotron Radiation Facilities (SSRF) for the sample preparation and measurement. This work was supported by the National Science Foundation of China Grant Nos. 51331004, 11374228 and 11205235, the National Basic Research Program of China under Grant No. 2015CB921501, and the Innovation Program of Shanghai Municipal Education Commission No. 14ZZ038. Z. G. Zhu is supported by Hundred Talents Program of The Chinese Academy of Sciences.
\end{acknowledgments}

%\newpage

\end{document}